\documentclass[aps,prl,twocolumn,superscriptaddress,amsmath,notitlepage]{revtex4-1}

\usepackage{amsmath,graphicx,bm,wasysym,color}
\usepackage{amsmath,amsfonts,amsthm,amssymb,amsbsy,bbold,mathrsfs}
\usepackage[caption=false]{subfig}

\newcommand{\nbm}[1]{ #1 }
\begin{document}
%%%%%   TITLE PAGE   %%%%%%
\title{Spin dynamics of counterrotating Kitaev spirals via duality}
%$\alpha$, $\beta$, $\gamma$ Li$_2$IrO$_3$}
\author{Itamar Kimchi}
\affiliation{Department of Physics, Massachusetts Institute of
Technology, Cambridge, MA 02139, USA}
\author{Radu Coldea}
\affiliation{Clarendon Laboratory, University of Oxford Physics
Department, Parks Road, Oxford OX1 3PU, United Kingdom}
\begin{abstract}
Incommensurate spiral order is a common occurrence in frustrated magnetic insulators. Typically, all magnetic moments rotate uniformly, through the same wavevector. However the honeycomb iridates family Li$_2$IrO$_3$ shows an incommensurate order where spirals on neighboring sublattices are counter-rotating, giving each moment a different local environment. Theoretically describing its spin dynamics has remained a challenge: the Kitaev interactions proposed to stabilize this state, which arise from strong spin-orbit effects, induce magnon umklapp scattering processes in spin-wave theory. Here we propose an approach via a (Klein) duality transformation into a conventional spiral of a frustrated Heisenberg model, allowing a direct derivation of the dynamical structure factor. We analyze both Kitaev and Dzyaloshinskii-Moriya based models, both of which can stabilize counterrotating spirals, but with different spin dynamics, and we propose experimental tests to identify the origin of counterrotation.
\end{abstract}

%%%%%   MAIN TEXT   %%%%%%
\maketitle

% Introduction.

Quantum spin liquid phases \cite{WenBook} have enjoyed renewed
attention in recent years, driven by candidate material platforms.
Possible experimental settings in magnetic insulators
\footnote{For a few recent reviews see
Refs.~\onlinecite{Balents2014, Kee2016, LucilleReview2016, Ng2016}.} include the layered kagome
systems, the nearly-metallic organics, as well as iridates
including the recently explored family of honeycomb iridates,
(Na/Li)$_2$IrO$_3$ and the related $\alpha$-RuCl$_3$,
distinguished by their significant spin-orbit coupling. Here
Ir$^{4+}$ (Ru$^{3+}$) hosts an effective $S{=}1/2$, observed to
order magnetically at low temperature. While Na$_2$IrO$_3$ and
$\alpha$-RuCl$_3$ show collinear zigzag antiferromagnetism
\cite{Gegenwart2010, Hill2011,Cao2012,Taylor2012,Gegenwart2012,
Kim2014a,Kim2015a,Baenitz2015,Johnson2015,Burch2015,Burch2016,Nagler2016},
the three structural polytypes of the lithium iridate,
$\alpha$,$\beta$,$\gamma$-Li$_2$IrO$_3$, all order into an
unconventional incommensurate magnetic phase, involving
counterrotating spirals
\cite{Analytis2014,Takagi2014,Coldea2014,Coldea2014a,Coldea2016}.

Recent experiments on $\beta$-Li$_2$IrO$_3$ under high pressures \cite{Takagi2014} as well as hydrogenated $\alpha$-Li$_2$IrO$_3$ \cite{Takayama2016} under ambient pressure found no evidence for magnetic long-range order at base temperatures, raising the interesting possibility of a transition into a long-sought Kitaev quantum spin liquid. Robustly identifying the properties of such a phase is experimentally rather challenging as the defining long-range
entanglement cannot be directly measured in a solid, and the
expected emergent fractionalized excitations are predicted to
produce only broad spectral features
\cite{Moessner2014,Moessner2015,Moessner2016,Moessner2016a,Balents2016,Kitaev3D}.
A possible route to quantify proximity to spin-liquid physics is
through a knowledge of the appropriate Hamiltonian in the
magnetically-ordered phase, whose properties could in principle be
more directly accessible experimentally. This requires detailed
predictions for characteristic signatures in the spin dynamics for
various Hamiltonians to be able to distinguish between competing
models.

% Figure.
\begin{figure}[b]
\includegraphics[width=\columnwidth]{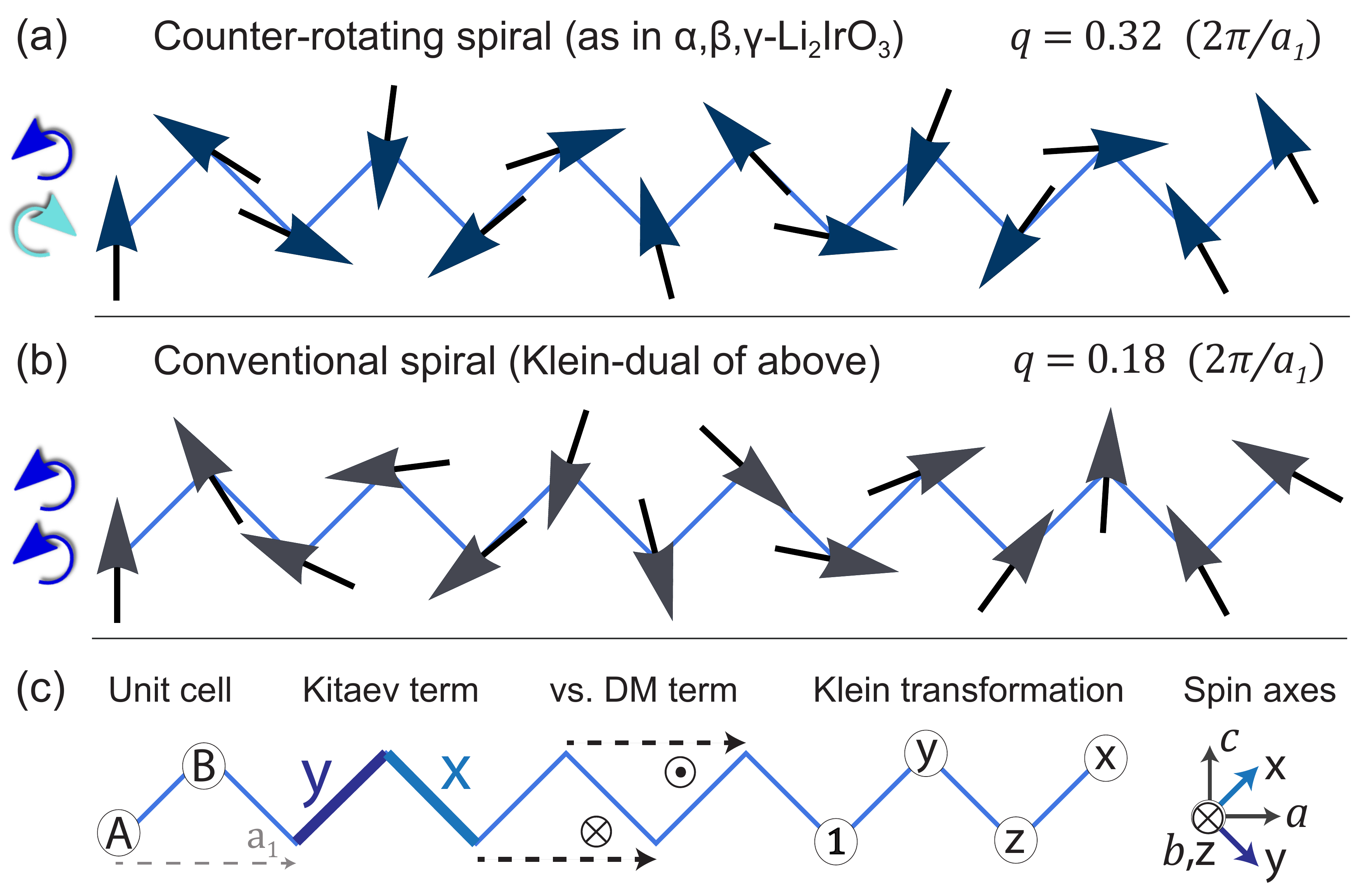}
\caption[]{ \textbf{Counterrotating spiral order of
$\alpha,\beta,\gamma$-Li$_2$IrO$_3$  as the Klein dual of a
conventional (co-rotating) spiral.} \\ \textbf{(a):} The
counterrotating spiral on a zigzag chain is the unifying common
feature of the magnetic structures of all three
$\alpha,\beta,\gamma$-Li$_2$IrO$_3$ honeycomb iridates. The bottom
sublattice rotates clockwise, while the top rotates
counterclockwise.
\\ \textbf{(b):}
The co-rotating spiral, of a conventional Heisenberg $J_1$-$J_2$
model, transforms by Klein duality into a counterrotating spiral
with a Kitaev-$J_1$-$J_2$ model and $\mathsf{xy}$ anisotropy.
\\ \textbf{(c):}
Competing models to stabilize counterrotation: Kitaev exchange
($\mathsf{x,y}$) or second-neighbor Dzyaloshinskii-Moriya (DM)
exchange (out/in for top/bottom bonds). The Klein transformation
$\mu\,{\in}\,\{ \mathsf{1,x,y,z}\}$ acts as identity $\mathsf{1}$
or by $\pi$ rotation around a spin's $\mathsf{x,y,z}$ axis. This
exact duality for the counterrotating spiral shows its stability 
and circumvents the magnetic umklapp of its Kitaev exchange for
computing its dynamical structure factor.} \label{fig:chain}
\end{figure}
% Figure.

The counterrotating spiral orders in
$\alpha,\beta,\gamma$-Li$_2$IrO$_3$ offer a promising avenue for
such an approach. However, theoretically computing the spin
dynamics has proven to be a nontrivial task. As we show below, the
barrier consists of strong magnon umklapp scattering, associated
both with the nonuniform spin environment of counterrotation as
well as with the lack of any continuous spin rotation symmetry in
the Hamiltonian. A similar issue was recently discussed for
$\beta$-CaCr$_2$O$_4$ \cite{Lake2015,Chapon2010}. Easy-axis and
easy-plane anisotropy, as well as antisymmetric
Dzyaloshinskii-Moriya (DM) exchange, which are expected to arise
from spin-orbit coupling, can preserve a continuous SO(2) symmetry
subgroup; in contrast, the ``Kitaev'' exchange of Kitaev's
honeycomb spin liquid \cite{Kitaev2006}, proposed to arise in the
honeycomb iridates
\cite{Khaliullin2009,Khaliullin2010,Kim2015,Kee2014,Kim2014,KHbhc}, breaks it
down to a discrete subgroup. Such a reduced symmetry in a minimal
Hamiltonian implies a remarkable spin-orbit coupling effect.

In this work we theoretically analyze the spin dynamics of a
minimal 1D model on a zigzag chain with coplanar $\mathsf{xy}$
spiral order with counterrotation on top/bottom sites as shown in
Fig.~\ref{fig:chain}(a). This captures the unifying common feature
of the magnetic structures in all three Li$_2$IrO$_3$ structural
polytypes; the actual structures differ in the value of the spin
rotation angle, the magnitude of the tilt of the rotation plane away from the
$\mathsf{xy}$ plane, and the pattern of those tilts between
adjacent chains, and we consider the tilts to be secondary features
left for future work. We describe the spin rotation along the
zigzag chain via a magnetic ordering wavevector $q$ in units of
$2\pi/a_1$, where $a_1=5.16$~\AA{} is the repeat distance along
the zigzag chain, see Fig.~\ref{fig:chain}(c). In this
description\cite{Coldea2014,Coldea2014a,Coldea2016} $q=0.32$ for
$\alpha$ and $0.28$ for $\beta$ and $\gamma$-Li$_2$IrO$_3$. 
It is important to note \cite{SuppMat}
that while for maximum generality and simplicity we
focus here on the parent 1D model, ultimately we want properties that are relevant for 3D systems.
Hence we are not interested in the true quantum excitations of an isolated 1D chain \cite{Kim2016},  which are usual 1D spinons. Instead, using the spin-wave method we expose precisely those features which are common to the 2D and 3D ordered materials. Our goal is to capture the ``semiclassical'' quantum fluctuations, appropriate for
the real materials, within a unified transparent setting.

The Hamiltonians we study are constructed as the Klein duals of
the known parent Hamiltonians for conventional spirals. The Klein
duality, a four-sublattice spin transformation whose
site-dependent $\pi$ rotations connect to the Kitaev exchange via
the multiplication rules of the Klein four group, was previously
used to expose a fluctuation-free point in a stripy
antiferromagnet \cite{Khaliullin2010} among other contexts
\cite{Okamoto2002,Okamoto2003,Khaliullin2005,Khaliullin2010,Khaliullin2012,KHbhc,Rachel2012,Li2015}.
Here we find that it transforms a co-rotating spiral in a
frustrated $J_1$-$J_2$ model into a counterrotating spiral in a
Kitaev-based model, with additional $J_2$ $\mathsf{xy}$ anisotropy
appropriate for the  $\mathsf{xy}$-coplanar spiral mode. We
compare this mechanism against a model of antisymmetric DM
couplings, here required to be purely intra-sublattice and with a
sublattice-dependent orientation \cite{Valenti2016}. We compute
the dynamical spin structure factor for various models of both
classes, through a rotating frame exposed by the duality
transformation. The dynamics in the Kitaev-based model are found
to be quite unusual, but can be interpreted via the duality to the
$J_1$-$J_2$ model's well-understood dynamics.

The general Hamiltonian consists of the following,
\begin{align}
\label{eq:H}
H&=\sum_{\langle i j\rangle} \left[
K S_i^{\gamma_{ij}}S_j^{\gamma_{ij}}
+ J_{1}^{\mathsf{xy}} (S_i^{\mathsf{x}}S_j^{\mathsf{x}}\,{+}\,S_i^{\mathsf{y}}S_j^{\mathsf{y}})
+ J_{1}^{\mathsf{z}} S_i^{\mathsf{z}}S_j^{\mathsf{z}} \right]
 \\
& +\sum_{\langle \langle i j\rangle \rangle} \left[
 J_{2}^{\mathsf{xy}} (S_i^{\mathsf{x}}S_j^{\mathsf{x}}{+}S_i^{\mathsf{y}}S_j^{\mathsf{y}})
+ J_{2}^{\mathsf{z}} S_i^{\mathsf{z}}S_j^{\mathsf{z}}
\pm D_2\, \hat{\mathsf{z}} {\cdot} \vec{S_i}{\times} \vec{S_j}
 \right]
  \nonumber
  \end{align}
where $\langle i j\rangle$ and $\langle \langle i j\rangle
\rangle$ refer to first and second neighbor bonds, respectively,
$\gamma_{ij}\in \{\mathsf{x,y} \} $ is the Kitaev bond type, and
the Dzyaloshinskii-Moriya coupling $D_2$ is oriented as in
Fig.~\ref{fig:chain}c), with $j{>}i$ and $\pm$ sign for the A/B
sublattice.

% Figure.
\begin{figure}[b]
\includegraphics[width=\columnwidth]{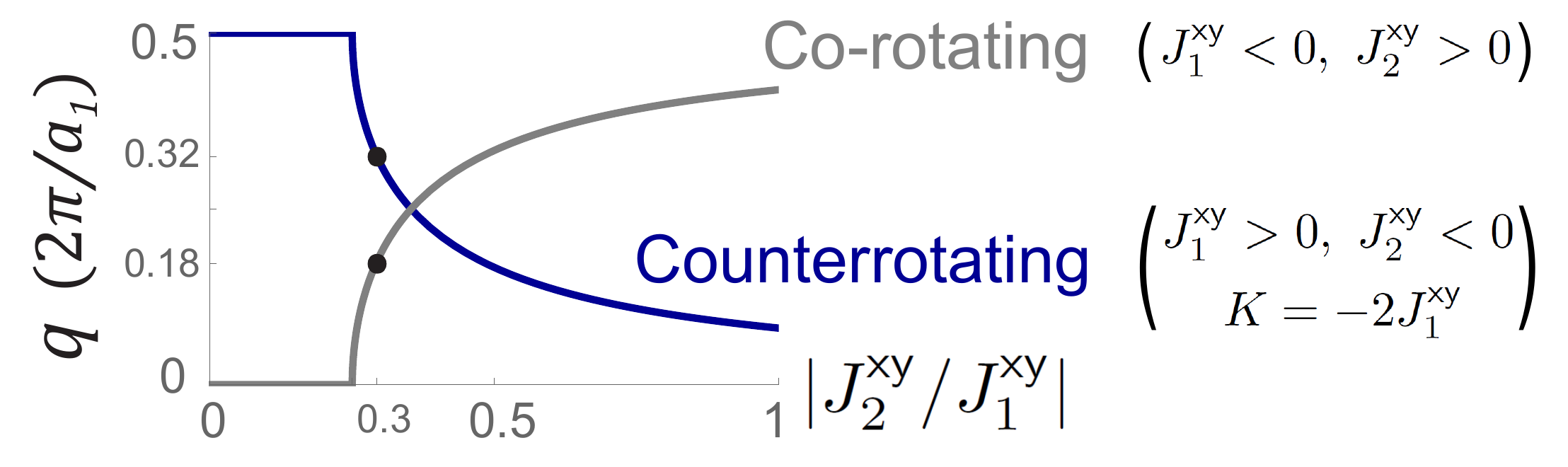}
\caption[]{ \textbf{Duality of classical spirals.} The classical
Heisenberg $J_1$-$J_2$ model with ferromagnetic $J_1<0$ and
frustrating second-neighbor $J_2>0$ has a (co-rotating) spiral
ground state with nonzero wavevector $q$ for $|J_2/J_1|>0.25$
(gray curve). With easy-plane $\mathsf{xy}$ anisotropy, the
resulting $\mathsf{xy}$-plane spiral is independent of
$J_{1}^{\mathsf{z}},J_{2}^{\mathsf{z}}$ and depends only on
$J_{2}^{\mathsf{xy}}/J_{1}^{\mathsf{xy}}$. The Klein
transformation produces a counterrotating spiral with
$q\rightarrow \pi/a_1-q$ (blue curve) while flipping the signs of
$J_{1}^{\mathsf{xy}},J_{2}^{\mathsf{xy}},J_{1}^{\mathsf{z}}$,
preserving $J_{2}^{\mathsf{z}}$ and creating a Kitaev exchange
$K={-}2J_{1}^{\mathsf{xy}}$. The resulting model has the
counterrotating spiral shown in Fig.~\ref{fig:chain}(a) as its
classical ground state. } \label{fig:wavevectors}
\end{figure}
% Figure.
% Figure.
\begin{figure*}[t]
\includegraphics[width=1.96\columnwidth]{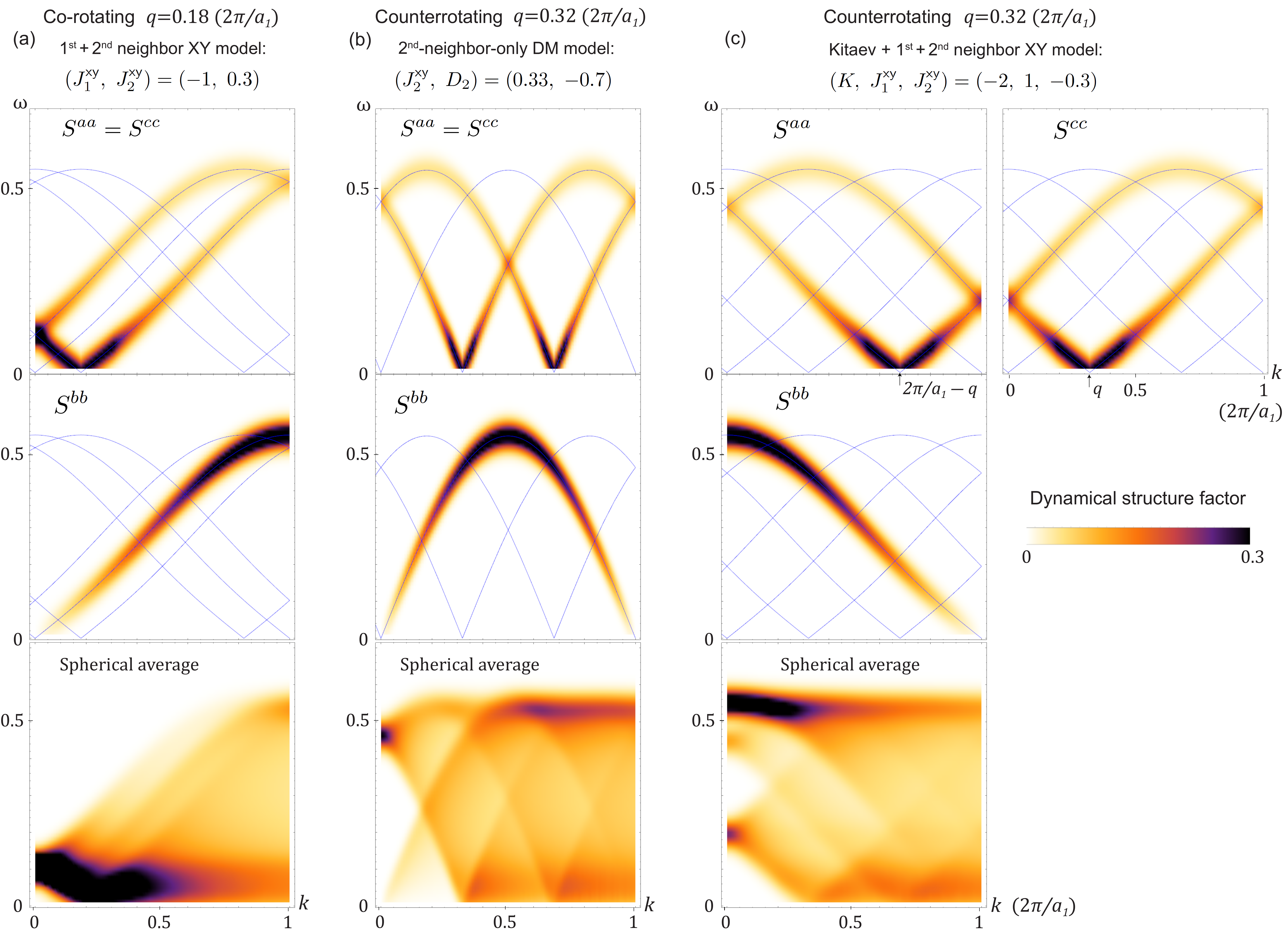}
\caption[]{\textbf{Dynamical structure factor signature of Kitaev
exchange.} The dynamical correlations of various spin
polarizations ($a,b,c$ axes defined in Fig.~\ref{fig:chain}c) are
computed via spin wave theory for two possible models of the
counterrotating spiral: decoupled sublattices with
pure-second-neighbor DM exchanges of opposite signs  (column b),
and nearest-neighbor Kitaev exchange together with smaller
easy-plane $J_1$-$J_2$ (column c). The plots shown were computed
for the minimal models with $J_{1}^{\mathsf{z}},J_{2}^{\mathsf{z}}
\rightarrow 0$. (Color is the dynamical spin structure factor,
convolved with a $\sigma{=}0.025$ energy Gaussian; thin blue lines
are underlying spin wave dispersions.) Magnetic umklapp
scattering, which usually breaks down spin waves of the Kitaev
exchange, was avoided by tuning to the duality with the
conventional co-rotating spiral of a $J_1$-$J_2$ XY model (panel
a). The Klein duality between panels (a) and (c)  shifts
wavevectors by $\pm \pi/a_1$ for $S^a,S^c$ and by $2\pi/a_1$ for
$S^b$, producing distinctive signatures for the Kitaev exchange;
for example, the shifted $S^{bb}$ is evident in the spherical
average, via the strong signal at high energy and low
momentum.}\label{fig:struct}
\end{figure*}
% Figure.

% Main paper.

\textbf{Classical ground states: mechanism for stability of the
counterrotating spiral.} First let us consider how to stabilize
the corotating and counterrotating spirals as ground states for
various terms in this Hamiltonian. There are two known mechanisms
for stabilizing conventional (corotating) spiral orders: (A)
Frustration from competing exchanges, such as ferromagnetic
nearest-neighbor and antiferromagnetic second-neighbor exchanges;
and (B) DM couplings. As an example of mechanism (A), we take a
$J_1$-$J_2$ ($K=0$) model with $J_1<0$ and $J_2>0$; its classical
ground state is a spiral order with a rotation angle between
consecutive sites $\arccos(-J_1/4J_2)$ for $J_2>|J_1|/4$. For
mechanism (B), the rotation angle is $\arctan(D/J_1)$ for the
usual nearest-neighbor DM model. When the zig-zag chain separates
into two decoupled A/B chains with DM interaction of opposite sign
the angle of rotation for each chain is $\theta_{A,B} = \pm
\arctan(D_2 / J_2)$.

The Klein duality, which maps a conventional co-rotating spiral to
a counterrotating spiral, transforms these conventional spiral
Hamiltonians to produce Hamiltonians for the counterrotating
spiral. It is easy to see (Fig.~\ref{fig:chain}) how the classical
conventional spiral order is transformed, by the rules of the
Klein transformation, into the counterrotating spiral order, with
$q \rightarrow \pi/a_1 -q$. Let us then consider how the
transformation acts on the Hamiltonians for mechanisms (A) and (B)
above, or relatedly on the Hamiltonian Eq.~\ref{eq:H} at $K=0$ and
uniform orientation of the DM term ($-D_2$ rather than $\pm D_2$).
It is easy to show the following action for the Klein
transformation:
\begin{align}
(J_{1}^{\mathsf{xy}},\ J_{2}^{\mathsf{xy}},\ J_{1}^{\mathsf{z}},\ J_{2}^{\mathsf{z}})\  &\leftrightarrow \ (-J_{1}^{\mathsf{xy}},\  -J_{2}^{\mathsf{xy}}, \  -J_{1}^{\mathsf{z}},\ +J_{2}^{\mathsf{z}})  \\
- D_2 \ &\leftrightarrow \ \pm D_2 \\
(K=0) \  &\leftrightarrow \   (K=-2 J_{1}^{\mathsf{xy}} )
\label{eq:duality}
\end{align}
A Kitaev term is produced, with twice the magnitude and opposite
sign relative to the $J_{1}^{\mathsf{xy}}$ term. This
transformation is a duality, i.e.\ it maps Eq.~\ref{eq:H} to
itself with a different set of parameters.

A known Hamiltonian for a conventional spiral thus produces a
Hamiltonian for the counterrotating spiral, via the mapping above.
The dual of mechanism (B) is obvious -- one can force
counter-rotation between sublattices by giving opposite signs to
pure-second-neighbor (intra-sublattice) DM terms, as in
Eq.~\ref{eq:H}. The dual of mechanism (A) however produces a
Kitaev-based model, with additional first and second neighbor
Heisenberg-type terms, whose classical ground state is the
counterrotating spiral. We note that the Klein duality necessarily
introduces easy-plane anisotropy via the differing transformation
of $J_{2}^{\mathsf{z}}$. Since the
$J_{1}^{\mathsf{z}},J_{2}^{\mathsf{z}}$ couplings do not change
the nature of the spiral order when the spin rotation plane is
$\mathsf{xy}$, i.e.\ for sufficient easy-plane $\mathsf{xy}$
anisotropy, a minimal description is afforded by setting
$J_{1}^{\mathsf{z}}{=}J_{2}^{\mathsf{z}}{=}0$. The result
(Fig.~\ref{fig:wavevectors}) is a
Kitaev-$J_{1}^{\mathsf{xy}}$-$J_{2}^{\mathsf{xy}}$ model whose
classical ground state is the counterrotating spiral.

\textbf{Spin dynamics and magnetic umklapp from spin-orbit
coupling.} To compute the dynamical structure factor via spin wave
theory, one transforms the Hamiltonian Eq.~\ref{eq:H} into a
``rotating'' (or ``moving'') frame, i.e.\ a site-varying
coordinate system which is locally aligned with the spin
orientation in the ordered spiral configuration. In the following
we find it convenient to use the orthorhombic axes $(a,b,c)$
instead of the Kitaev $(\mathsf{x,y,z})$ axes for the spin
components, with the relation \cite{UnifiedTheory}
$\nbm{\hat{\mathsf{x}}}=(\nbm{\hat{a}}+\nbm{\hat{c}})/\sqrt{2}$,
$\nbm{\hat{\mathsf{y}}}=(\nbm{\hat{a}}-\nbm{\hat{c}})/\sqrt{2}$
and $\nbm{\hat{\mathsf{z}}}=\nbm{\hat{b}}$ shown in
Fig.~\ref{fig:chain}(c), where $\nbm{\hat{\mathsf{x}}}$ indicates
a unit vector along $\mathsf{x}$ and so on.
Let $R[\theta]$ be a rotation by angle $\theta$ around the spin
$\mathsf{z}\equiv b$ axis. The local spin
orientation in the wavevector-$q$ spiral is expressed by
$e^3
\equiv R[{-}\eta_s q r] \cdot \hat{c}$.
Here the sublattice sign
$\eta_s$ is $\eta_s=\mp$ on the A/B sublattice for the
counterrotating spiral, or is uniformly $\eta_s=+$ for the
co-rotating spiral. The local coordinate system $e^{\pm}\equiv
R[{-}\eta_s q r] \cdot (\hat{a}\pm i \hat{b})$ can then be used
to write the spin operator as $\vec{S}=e^3 S^3+(e^- S^+ + e^+
S^-)/2$. In the $1/S$ spin wave expansion, $S^3\rightarrow
1/2-b^\dagger b$ and $S^{\pm} \rightarrow b,b^\dagger \equiv a^\pm
$.
The spin wave Hamiltonian is then
$H_{\text{SW}} = \sum_{i j} \left[ \tilde{J}_{i,j}^{\mu \rho} \,
\sigma^1_{\rho \nu}/8
 - \delta_{\mu \nu} \delta_{i j} E_{\text{cl}}/2
  \right]\left(a^\mu_i \right)^\dagger a^\nu_j$ with repeated indices summed.
   The important ingredient is the interaction matrix in the rotating frame,
$\tilde{J}_{i,j}^{\mu,\nu} \equiv e^\mu_i {\cdot} J_{i,j} {\cdot} e^\nu_j $, where $J_{i,j}$ is the spin interaction matrix between spins $i,j$ associated
with Eq.~\ref{eq:H}.

We thus turn to evaluate the interactions in the rotating frame, $\tilde{J}_{i,j}^{\mu \nu}$. The rotation
around $\mathsf{z} \equiv  b$ leaves $\hat{b}$ invariant,
$e^\pm = (R \cdot \hat{a}) \pm i\hat{b}$, so its effects are contained in  the $R \cdot \hat{a}$ component; for
concreteness, we can isolate it by setting $J_{1}^{\mathsf{z}}=J_{2}^{\mathsf{z}}=0$, in which case $\tilde{J}  \rightarrow \hat{a} \cdot R^T \cdot J  \cdot R \cdot \hat{a}  $.
Evaluating this term on nearest-neighbor bonds $(i,j)$, which connect
opposite sublattices, we find\cite{SuppMat}
\begin{align}
\label{eq:rotated}
\tilde{J}_{i,j} &= \hat{a} \cdot R^T[-\eta_s q r] \cdot J_{i,j} \cdot R[\eta_s q (r+a_1/2)] \cdot \hat{a}  \\
&=
-\frac{K}{2}\sin\left(\frac{q a_1}{2}\right)
+ \left[ J_{1}^{\mathsf{xy}}+\frac{K}{2}\right] \cos \left(\frac{q a_1}{2}+ 2 q r\right)
 \nonumber
\end{align}
where $\eta_s=\mp$ is defined by the A/B sublattice of site $i$,
at position $r$. The explicit dependence on coordinate $r$ in the last term --- the
rotated Hamiltonian is not translationally invariant ---  changes
the spin wave physics drastically. This is exposed by Fourier
transform, where the expression above produces magnetic umklapp
terms such as $b^\dagger_{k} b_{k+2q}$. The magnons experience
magnetic umklapp scattering that changes their wavevector by
multiples of $q$. Even if $q$ is taken to be approximately
commensurate, the wavevector quantum number $k$ is lost outside of
a highly-folded magnetic Brillouin zone; for incommensurate $q$,
the magnon wavevector $k$ becomes ill-defined.

One might generally expect to lose the wavevector quantum number
$k$ when translation symmetry is fully broken by an incommensurate
order; this is masked in conventional spirals through a rotating
frame, which relies on continuous SO(2) rotation symmetry in the
model Hamiltonian. The SO(2)-symmetric $J_{2}^{\mathsf{xy}}$-$D_2$
second-neighbor model of the counterrotating spiral can similarly
preserve the magnon wavevector $k$. However the counterrotation
configuration means that each spin has a different local
(nearest-neighbor) environment, giving rise to magnetic umklapp
processes even through the SO(2)-symmetric $J_{1}^{\mathsf{xy}}$
term, as well as through the discrete-symmetry $K$ terms. The loss
of $k$ as a good quantum number is fully apparent.

Here we circumvent the magnetic umklapp scattering by tuning
parameters to the duality with the co-rotating spiral. Recall from
Eq.~\ref{eq:duality} that the counterrotating spiral Hamiltonian
with $K=-2J_{1}^{\mathsf{xy}}$ is dual to a $J_1$-$J_2$ XY model.
The continuous SO(2) symmetry group of the XY model is preserved
in an altered form by the duality, allowing the Hamiltonian at
$K=-2J_{1}^{\mathsf{xy}}$ to preserve the magnon quantum numbers.
Indeed, Eq.~\ref{eq:rotated} shows that the translation symmetry
in the rotated frame is restored when $K+2J_{1}^{\mathsf{xy}}=0$.
We proceed by analyzing this case. Perturbations away from this
parameter point will generically open gaps in the spin wave
dispersions via Bragg reflections through multiples of the spiral
wavevector $q$, such as at wavevectors $k=\pm q$, as well as mix
the $S^a$, $S^c$ spin polarizations.

Using the counterrotating spiral model produced by the duality,
the dynamical structure factor can be computed straightforwardly
by diagonalizing the spin wave Hamiltonian. The results are shown
in Fig.~\ref{fig:struct}, for various polarizations as well as for
a spherical average relevant to powder samples. The Kitaev-based
model shows unusual features, which are nevertheless transparently
related, via the Klein duality, to the usual features from the
conventional spiral. The duality shifts magnon wavevectors by $\pm
\pi/a_1$ for the $S^a$ and $S^c$ spin components, respectively,
and by $2\pi/a_1$ (corresponding to N\'{e}el correlations) for the
$S^b$ spin component. The Bragg peaks and intensity pattern are
thus found by appropriately shifting the known structure factor of
the $J_1$-$J_2$ conventional spiral. Observe that the
counterrotating spiral can be considered as a sum of two distinct
$S^a$,  $S^c$ spin density waves $\pi/2$ out-of-phase. The two
sublattices have in-phase $S^c$ but $\pi$-out-of-phase
($2\pi/a_1$-modulated) $S^a$, producing $S^c$-polarized Bragg
peaks at $k=0\pm q$, but $S^a$-polarized Bragg peaks at
$k=2\pi/a_1 \pm q$. Universal, linearly-dispersing Goldstone modes
with the same polarization as the Bragg peaks emerge from $q$ and
$2\pi/a_1 \pm q$ positions. The $S^{bb}$ dynamical correlations
(out-of-plane fluctuations) contain a mode with maximum energy and
strong intensity at the zone center ($k=0$), as in other
Kitaev-based models \cite{Khaliullin2012}. We expect these generic
features survive when the 1D chains are coupled together
\cite{UnifiedTheory} as in the actual 2D and 3D honeycomb iridates
and to help distinguish between Kitaev or other exchange models.

\textbf{Conclusion.} We have identified a transparent theoretical
mechanism for the key feature in the unconventional magnetic
orders recently observed in three honeycomb iridates. These
materials host different crystal structures but nonetheless their
magnetism shares the unifying feature of counterrotating spirals,
with opposite handedness in neighboring sublattices. This magnetic
configuration, as well as a Kitaev-based parent Hamiltonian, are
constructed by acting with the Klein duality on the
well-understood frustrated $J_1$-$J_2$ model of a spiral order.
This connection also enables us to solve for the spin dynamics in
this system, and to interpret them transparently. We have
identified key features in the dynamical structure factor, that
could be tested via polarized and unpolarized inelastic neutron
scattering or resonant inelastic x-ray scattering experiments. Our
work helps build towards a full understanding of the lattice-scale
model Hamiltonians for these systems, which would shed further
light on the unusually similar features across these disparate
materials, as well as enable a controlled identification and
understanding of possible proximity to a spin liquid state.

\textbf{Acknowledgements.} We thank Ashvin Vishwanath, Lucile
Savary, Samuel Lederer, Jonathan Ruhman, and Yong-Baek Kim for
related discussions. I.K.\ was supported by the MIT Pappalardo
postdoctoral fellowship. R.C.\ acknowledges support from EPSRC (U.K.)
through Grant No. EP/H014934/1.

%%%%%%   Supplementary Material   %%%%%%
%
%%\clearpage
\section*{Supplementary Material}

The Supplementary Material is divided into five sections. Section
(1) shows additional plots of the structure factor. Section (2)
gives a short note on the 1D and spin wave approximations. Section
(3) gives further detail on the duality between conventional
co-rotating and the unconventional counterrotating spirals.
Section (4) gives further detail on the spin interactions in the
spiral rotating frame, and the spin wave Hamiltonian. Section (5)
gives further detail on the computations of the dynamical spin
structure factors.

\subsection{Supplementary Material: (1) Additional plots}
\begin{figure}[t]
\includegraphics[width=\columnwidth]{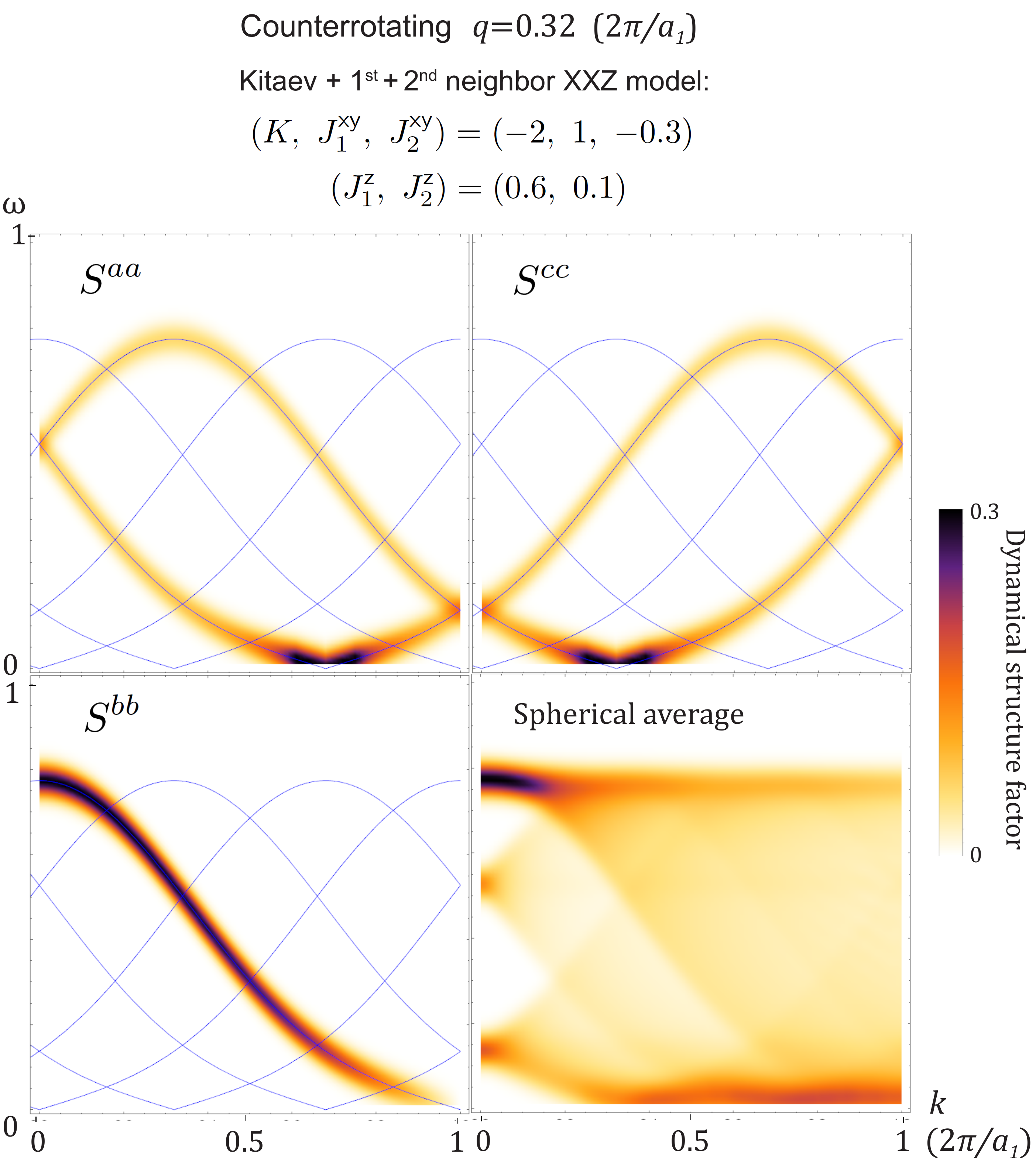}
\caption[]{ \textbf{Dynamical structure factor signature of Kitaev
exchange -- plots for additional parameters.} Additional structure
factor plots for a model with nearest-neighbor Kitaev exchange
together with smaller easy-plane $J_1$-$J_2$, with intermediate
values for the easy-plane anisotropies. (Color intensity is
dynamical spin structure factor, convolved with $\sigma{=}0.025$
energy Gaussian; thin blue lines are underlying spin wave
dispersions.) Though the spin wave dispersions are modified, the
qualitative nature of the distinctive feature remains. }
\label{fig:appendixstruct}
\end{figure}

\begin{figure}[t]
\includegraphics[width=\columnwidth]{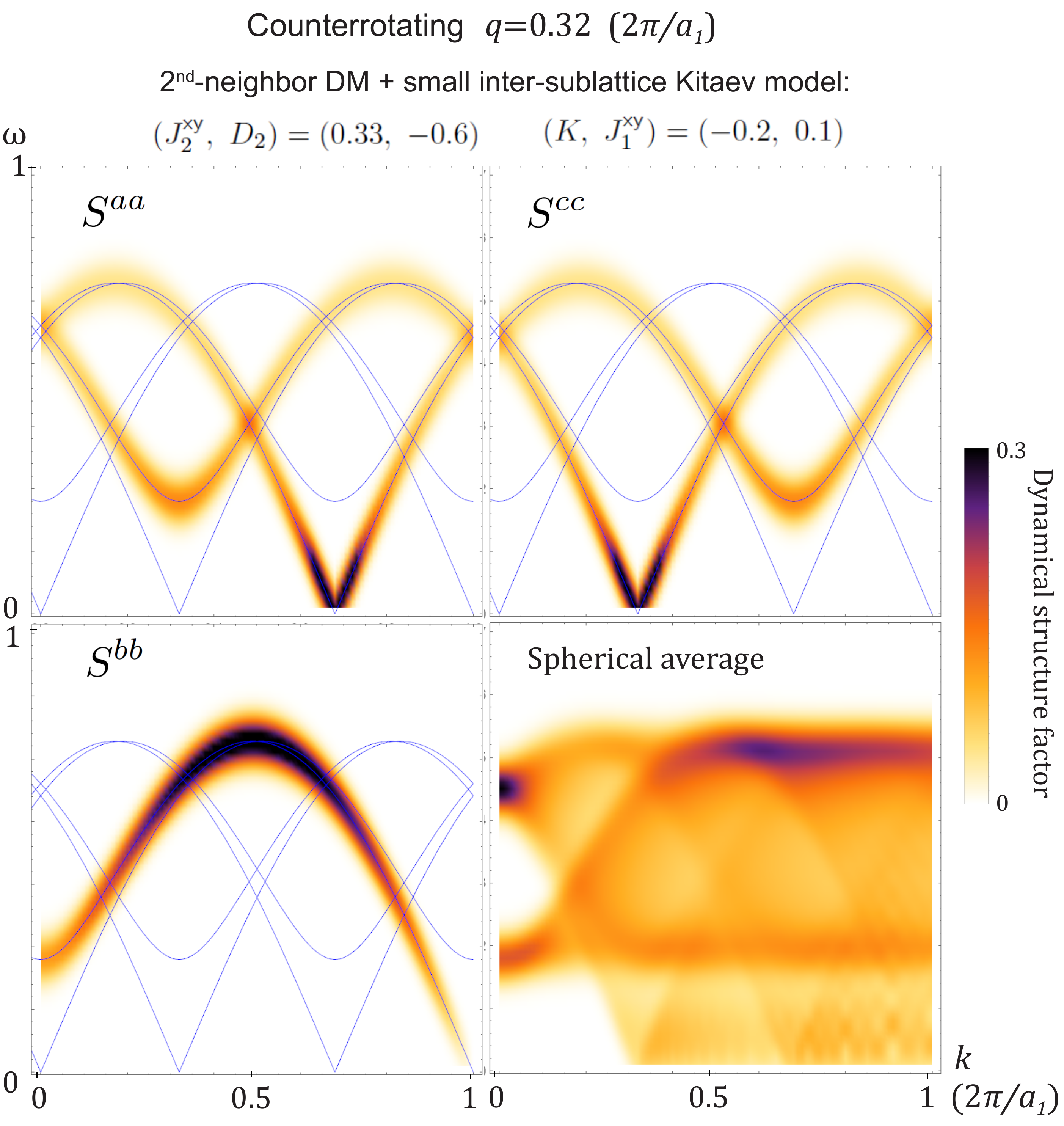}
\caption[]{\textbf{Dynamical structure factor for second-neighbor
DM model with additional small first neighbor Kitaev-based
exchange.} Additional structure factor plots for a model based on
the DM mechanism but with an additional perturbation. The
second-neighbor DM model whose plots are shown in Fig.~3 panel (b)
consists of two independent chains, the A sublattice and the B
sublattice, which are fully decoupled --- each sublattice has its
own usual Heisenberg plus DM model, though the sublattices have
opposite DM vectors. The relative phase between the two spirals is
then chosen spontaneously, via spontaneous symmetry breaking of
the SO(2) symmetry. Small inter-sublattice couplings will
explicitly modify the ground state, but will not drastically
modify the structure factor at finite energies. This can be seen
explicitly by adding small coupling between the two sublattices,
in the form of the Kitaev model of Fig.~3 panel (c), which fix the
ground state into the Kitaev counterrotating spiral of Fig.~1
panel (a). The structure factor of the resulting model is shown
here. (Color is dynamical spin structure factor, convolved with a
$\sigma{=}0.025$ energy Gaussian; thin blue lines are underlying
spin wave dispersions.) The $S^{aa}$ and $S^{cc}$ structure
factors are no longer equal due to the presence of the Kitaev
exchange. The inter-sublattice Kitaev coupling gaps out the
dispersions at wavevector $q$ in $S^{aa}$ and $2\pi/a_1-q$ in
$S^{cc}$, but as long as it is not too large, it leaves the strong
intensity region of $S^{bb}$ essentially unaffected. }
\label{fig:appendixstruct2}
\end{figure}

See Figs.~\ref{fig:appendixstruct} and ~\ref{fig:appendixstruct2}
for plots of the structure factor for two additional models: (A) a
Kitaev-based model with nonzero values of ${\mathsf z}$-axis
couplings, away from the XY limit, in
Fig.~\ref{fig:appendixstruct}; and (B) a second neighbor DM model
perturbed by adding small inter-sublattice couplings, in
Figs.~\ref{fig:appendixstruct2}. The results for these models and
for various other intermediate parameters maintain the features
described in the main text.

\subsection{Supplementary Material: (2)
A note on quantum vs.\ classical approximations in the 1D minimal
model} We note that while for maximum generality and simplicity we
focus here on a 1D model, ultimately we want properties that are
relevant for 3D systems. Hence we are not interested in the true
quantum excitation spectrum of an isolated 1D chain, as would be
accessible within e.g. DMRG. Such a spectrum would contain the
usual 1D spinon excitations, which do not generalize to 2D and 3D
magnetic systems. For example, the quantum ground state of the XXZ
$J_1$-$J_2$ model is known, and has short ranged spiral order
instead of a long range spiral state, as is necessary due to the
Mermin-Wagner theorem in 1D since the incommensurate spiral breaks a continuous symmetry. 
By treating the 1D minimal model within a spin-wave
approximation, we avoid any purely-1D effects, such as spin-charge separation (which here would be manifested as spurious 1D deconfined spinons) and Mermin-Wagner
destruction of long range order, which are not relevant for the
actual materials. 
Since our 1D system serves as a
minimal model for full 3D systems, which are known to exhibit long
range ordering, the appropriate minimal 1D model --- that captures
the essential low-energy physics of the 3D quantum model relevant
for the real materials --- is the 1D classical, rather than
quantum, model, together with its semiclassical spin wave
spectrum.

It is also worth noting a sublety in applying spin-wave theory to materials which, though exhibiting an ordered ground state, are thought to be proximate to a quantum spin liquid.  Proximity to a spin liquid cannot be established in the 3D quantum phase diagrams, but a pure-Kitaev Hamiltonian is exactly solvable and exhibits a spin liquid ground state in both 2D and 3D. A model where the Kitaev exchange is significantly larger than any other exchange is then, in some loosely defined sense, potentially proximate to the large-Kitaev quantum spin liquid. Such proximity might suggest that quantum fluctuations are too strong for spin wave theory. We avoid this issue by studying a model where $K$ is larger than any other term, but which nevertheless is exactly dual to a pure-Heisenberg $J_1$-$J_2$ model. The $J_1$-$J_2$ model has quantum fluctuations, but has been well studied and its low energy excitations are thought to be captured well by a spin wave approach.

\subsection{Supplementary Material: (3) Classical solutions across the duality.}
The Klein duality transforms the co-rotating spiral into the counter-rotating spiral, and vice versa, as well as exchanging wavevector $q \rightarrow 2\pi/a_1-q$. Here we consider these two spiral orders related by the duality in detail. In the counterrotating spiral, the spin configuration is described by the following functions, on sublattice $A$ and sublattice $B$ separately:
\begin{align}
\vec{S}_{A,r} &=  \cos( q r) \hat{c} + \sin(q r) \hat{a}    \nonumber\\
\vec{S}_{B,r} &=  \cos( q r) \hat{c} - \sin(q r) \hat{a}
\end{align}
In the co-rotating spiral, the spin configuration is described by the same function on the two sublattices, i.e. the spin moment at a general site r (on either the A or B sublattice) is given by 
\begin{align}
\vec{S}_{r} =   \cos( q r) \hat{c} - \sin(q r) \hat{a}  
\end{align}

Let us consider the models with $D_2=0$, related by the duality.
The classical energies per site of the co-rotating and
counterrotating spiral models, $E_H$ and $E_K$ respectively, are
given by the following (in units of $S^2$):
\begin{align}
&E_H = J_{1}^{\mathsf{xy}} \cos(q a_1 /2) + J_{2}^{\mathsf{xy}} \cos(q a_1)
 \\
&E_K = (K/2) \sin(q a_1 /2) + J_{2}^{\mathsf{xy}} \cos(q a_1)
\label{eq:energies}
\end{align}
When parameters are taken to match under the duality relations
above, the two ground state energies become identical upon the
substitution $q\rightarrow \pi/a_1-q$.

On both sides of the duality, the incommensurate spiral with
nonzero wavevector becomes the stable classical ground state for
$|J_{2}^{\mathsf{xy}}|>|J_{1}^{\mathsf{xy}}|/4$. This is well
known on the Heisenberg side, where this is the critical value of
AF $J_2$ needed to frustrate the FM $J_1$ order. Then the spiral
wavevector, determined by minimizing the energy, is as follows:
\begin{align}
&q_{\text{co-rot}} = \frac{2}{a_1}\arccos\left( \frac{|J_{1}^{\mathsf{xy}}|}{4 J_{2}^{\mathsf{xy}}}\right)
 \\
&q_{\text{counter-rot}} = \frac{2}{a_1}\arcsin\left( \frac{K}{8 |J_{2}^{\mathsf{xy}}|}\right)
\label{eq:wavevectors}
\end{align}
The DM-based model is straightforward; to study the effects of the
duality on $D_2$, note that $q\rightarrow \pi/a_1-q$ changes the
sign of the relevant quantity $\tan(q a_1)$.

The spiral ground state of the $J_1$-$J_2$ Heisenberg model spontaneously chooses a $+q$ or $-q$ ground state, as can be seen in the $q \rightarrow -q$ symmetry of $E_H$ in Eq.~\ref{eq:energies}. When the duality mapping is exact, this feature is preserved, though in an unusual form. The energy of the counterrotating spiral in this model, $E_K$ in Eq.~\ref{eq:energies}, also has two solutions: wavevector $q$, as well as wavevector $2\pi/a_1 - q$. The latter solution is the counter-rotating dual of the $-q$ co-rotating spiral. Both solutions can also be seen as allowed wavevector solutions of the right-hand-side of Eq.~\ref{eq:wavevectors}. The $ q \rightarrow (2\pi/a_1 - q)$ transformation can be enacted visually by flipping spins on one sublattice (equivalent to adding $2\pi/a_1$ to $q$) followed by a flip of the sense of rotation on both sublattices. In other words, the sense of the rotation in the counterrotating spiral depends on the sublattice, on the sign of the Kitaev exchange, and on whether the chosen solution involves  $S^c$ or  $S^a$ as the spin component which is aligned across the two sublattices. This manifestation of the $\pm q$ symmetry is a useful consistency check for the duality on these 1D models; however, it is not a robust feature on the Kitaev side of the duality, and disappears upon inclusion of inter-chain $I_c$ couplings which are expected to arise in certain models\cite{UnifiedTheory}. These energetically favor $S^c$ alignment, which is equivalent to explicitly choosing one of the two solutions as the unique ground state, namely wavevector $q$ in the notation above. 

\subsection{Supplementary Material: (4) Details of the spin wave Hamiltonian computation.}
First let us recall the definition of the local coordinate system.
We here write vectors in spin space using the orthorhombic axes
basis $(a,b,c)$ that allows a common description of the crystal
structure of all three polytypes of Li$_2$IrO$_3$. The Kitaev axes
are given by $\hat{\mathsf{x}},\hat{\mathsf{y}}=(\hat{a} \pm
\hat{c})/\sqrt{2}$, $\hat{\mathsf{z}}=\hat{b}$. The zigzag chains
are oriented along the diagonals of the orthorhombic structural
cell, with $\vec{a}_1=(\vec{a}\pm\vec{b})/2$, making an angle
$\sim$55$^{\circ}$ with the ($ac$) spin rotation plane
\cite{UnifiedTheory}. Let site $j$ be defined by its spatial
position $r$ and unit cell index $s$. The local coordinate system
is written as
\begin{align}
\label{eq:localz}
e^3_{r,s} &= -\eta_s \sin(q r) \hat{a} + \cos( q r) \hat{c}  \nonumber\\
&= \left\{  -\eta_s \sin(q r) ,\ 0,\  \cos( q r) \right\}
\nonumber\\
e_j^\pm &= \left\{\cos( q r),\ \pm i, \  \eta_s \sin(q r) \right\}
\end{align}
where $\eta_s$ is defined as follows.  For the Kitaev model
counterrotating spiral, $\eta_s = +1$ for sublattice $s=B$ and
$\eta_s=-1$ for sublattice $s=A$. For the Heisenberg model
co-rotating spiral, $\eta_s=+1$ independent of sublattice.

The spin wave Hamiltonian in real space is computed as follows. A useful
intermediate step is the following,
\begin{align}
H = \frac{1}{2} \sum_{i j} \bigg[ & \
(e^3_j {\cdot} \tilde{J}_{i j} {\cdot} e^3_{i}) \big(S^2 - S (b^\dagger_j b_j + b_j  b^\dagger_j -1)\big) \nonumber \\&
+ \frac{S}{2} \big(e_j^- b_j +e_j^+  b^\dagger_j  \big) {\cdot}  \tilde{J}_{i j} {\cdot} \big(e_{i}^-   b_{i}  +e_{i}^+ b^\dagger_{i} \big)  \bigg]
\end{align}
Note that the apparent symmetry between $b$ operators left and
right of $\tilde{J}$ results in the $\sigma^1$ term when using the
$a^\pm$ notation. The spin wave Hamiltonian is then found within the form shown in the main text,
\begin{align}
H_{\text{SW}} &=  \frac{1}{8} \sum_{i,j} \left[
e^\mu_i {\cdot} J_{i,j} {\cdot} e^\rho_j \sigma^1_{\rho \nu}
 -4 E_{\text{cl}} \delta_{\mu \nu} \delta_{i j}
  \right]\left(a^\mu_i\right)^\dagger a^\nu_j
  \label{eq:HSW}
\end{align}
where $J_{i,j}$ is the spin interaction matrix associated
with Eq.~1, taken between spins $i,j$; $\sigma^1$ is a
Pauli matrix; and summation over repeated $\mu,\nu,\rho$ indices
is implied, taking values $\pm 1$. We have subtracted a total energy shift $S(S{+}1)N
E_{\text{cl}}$; here $E_{\text{cl}}\equiv (1/2N_{\text{sites}})
\sum_{i,j} e^3_i \cdot J_{i,j} \cdot  e^3_j$ is the classical
energy.  Recall that  in the $1/S$ spin wave expansion, $S^3\rightarrow
1/2-b^\dagger b$ and $S^{\pm} \rightarrow b,b^\dagger \equiv a^\pm
$  where the $a^\pm$ spin wave operators are introduced for
convenience.

For completeness we record the full spin interactions in the
rotating frame. We write the interaction matrix in the $2\times 2$
subspace of the $\hat{a},\hat{c}$ basis, using Pauli matrices. Let
us use a notation where we list the coefficients of the Identity
matrix followed by the three Pauli matrices, i.e.\
$\sigma^{0,1,2,3}$ respectively. In this notation, the nearest
neighbor lab-frame spin interactions are $(J_1^\mathsf{xy} + K/2, \eta_s K/2,
0, 0)$, while the second neighbor (intra-sublattice) lab-frame
spin interactions are $(J_2^\mathsf{xy},0, i \eta_s D_2, 0)$.  The rotating frame spin interaction
matrices are then as follows. For nearest-neighbor bonds in the
counterrotating spiral,
\begin{align}
\tilde{J_1} \rightarrow
\Bigg(&\frac{K+2 J_1^\mathsf{xy}}{2} \cos \left(q \frac{4 r+a_1}{2}\right),\ \frac{1}{2} \eta_s  K \cos \left(\frac{q a_1}{2}\right),
\nonumber\\ & i \eta_s  \frac{K+2 J_1^\mathsf{xy}}{2}  \sin \left(q \frac{4 r+a_1}{2}\right),\ -\frac{K}{2}\sin \left(\frac{q a_1}{2}\right)\Bigg)
\end{align}
For nearest-neighbor bonds in the co-rotating spiral,
\begin{align}
\tilde{J_1} \rightarrow
\Bigg(&\frac{K+2 J_1^\mathsf{xy}}{2} \cos \left( \frac{q a_1}{2}\right),\ \frac{1}{2} \eta_s  K \cos \left(q \frac{4 r+a_1}{2}\right),
\nonumber\\ & -i \eta_s  \frac{K+2 J_1^\mathsf{xy}}{2}  \sin \left( \frac{q a_1}{2}\right),\ \frac{K}{2}\sin \left(q \frac{4 r+a_1}{2}\right)\Bigg)
\end{align}
For second-neighbor bonds, which lie within a single sublattice, for either co- or counter-rotation,
\begin{align} \tilde{J_2} \rightarrow
\bigg(&J_2^\mathsf{xy} \cos(q a_1) + D_2 \sin(q a_1),\ 0, \nonumber\\  &i \eta_s (D_2 \cos(q a_1) -J_2^\mathsf{xy} \sin(q a_1)),\  0
\bigg)
\end{align}
In each of these expressions, the diagonal matrix elements  of
$\tilde{J}$ with $\hat{a}$ ($\hat{c}$ ) can be read off as the sum
(difference) of the identity and $\sigma^3$ coefficients, ie the
first component plus (minus) the last component.

The momentum space spin wave Hamiltonian can be expressed as follows,
\begin{align}
& H_\text{SW}  = \frac{S}{4} \sum_{k, s s' \mu \mu'}  \ \tilde{H}^{\mu \mu'}_{s s'}(k)\   (a_{k, s}^\mu)^\dagger\ a_{k, s'}^{\mu'}
\end{align}
where
  \begin{align}\tilde{H}^{\mu \mu'}_{s s'}&
  =
  -4 (\hat{c} \cdot (\tilde{J_1} + \tilde{J_2}) \cdot \hat{c}) \nu^0_{\mu \mu'} \tau^0_{s s'} \\
  &+ 2  (\hat{a} \cdot (\tilde{J_1}C_1\tau^1_{s s'} + \tilde{J_2}C_2\tau^0_{s s'}) \cdot \hat{a} )(\nu^0_{\mu \mu'} + \nu^1_{\mu \mu'} )\nonumber\\
  &+ 2   (J_{1}^{\mathsf{z}}C_1\tau^1_{s s'} + J_{2}^{\mathsf{z}}C_2\tau^0_{s s'}) (\nu^0_{\mu \mu'} - \nu^1_{\mu \mu'} )
  \nonumber\\
  C_\alpha &\equiv \cos\left(\frac{\alpha k a_1}{2}\right)
  \nonumber
  \end{align}
 where $\nu$ and $\tau$ are Pauli matrices for the Bogoliubov $\mu$ and sublattice $s$ indices. The diagonal matrix elements of $\tilde{J}$ can be read off from the equations in the paragraph above, and  $\tilde{J_1}$ and $\tilde{J_2}$ are the spin interactions between first and second neighbors respectively.

We may also explicitly record the matrix $\tilde{H}^{\mu \mu'}_{s
s'}(k)$, specialized to the case of most interest, $D_2=0$. We
define
\begin{align}
&L_1 = \left(\frac{K}{2}+J_{1}^{\mathsf{xy}}
\right)\cos\left(\frac{q a_1}{2}\right) - \frac{K}{2}
\sin\left(\frac{q a_1}{2}\right)
\\
&L_2=J_{2}^{\mathsf{xy}} \cos( q a_1)
 \end{align}
With these expressions, the Hamiltonian matrix may be written as
 \begin{align}
\tilde{H} = A \nu^0 \tau^0 + B \nu^0 \tau^1 +C \nu^1 \tau^0 + D \nu^1 \tau^1
 \end{align}
with $A,B,C,D$ given by
 \begin{align}
& A= -4  \left( L_2 + \zeta  L_1\right) + 2  \cos(k a_1)
(L_2+J_{2z}) \nonumber \\& B= 2 \cos\left(\frac{k a_1}{2}\right)
(L_1+J_{z}) \nonumber \\& C= 2 \cos(k a_1)  (L_2-J_{2z}) \nonumber
\\& D = 2 \cos\left(\frac{k a_1}{2}\right)  (L_1-J_{z})
 \end{align}

The $\zeta = \pm$ sign corresponds to the Heisenberg/Kitaev cases,
taking the value $\zeta =+1$ for the co-rotating spiral, and
$\zeta =-1$ for the counter-rotating spiral. The sublattice
indices $s,s'$ (of the matrices $\tau$) and the Bogoliubov indices
$\mu,\mu'$ (of the matrices $\nu$) are suppressed.

\subsection{Supplementary Material: (5) Details of the dynamical structure factor computation.}

Now we turn to computing the dynamical spin correlators. We find
that the structure factor is diagonal in the $a,b,c$ basis for the
spin axes, rather than the Kitaev $\mathsf{x,y,z}$ axes; only in
the $a,b,c$ axes do all off-diagonal terms cancel.

The dynamical spin
structure factor is expressed by the following,
\begin{align}
& S^{a a}(p,\omega)
= \frac{S}{32}\sum_{k=\pm q} F_{(+,+)}(p+k,\omega)
\nonumber
\\ &
S^{c c}(p,\omega)
= \frac{S}{32}\sum_{k=\pm q} F_{(+,\zeta)}(p+k,\omega)\nonumber
\\ &
S^{b b}(p,\omega)
= \frac{S}{8} F_{(-,+)}(p,\omega)
\end{align}
The differing second argument for $F$ in $S^{cc}$ for the counterrotating case arises from the product of sublattice signs $\eta_s \eta_{s'}$, which arises in this case through the axis perpendicular to the local spin orientation.
The function $F_{(\eta_m,\eta_t)}(k,\omega)$, where $\eta_m$ and $\eta_t$ are $\pm$ signs, is defined by
\begin{align}
& F_{(\eta_m,\eta_t)}(k,\omega) = \sum_{n=1,2}\delta(\omega-\omega_n(k)) \times \\
&\bigg[\sum_{ss'\mu\mu'} \left( T^{(-1,\mu)}_{(n,s)}(k)\right)^\dagger
 \nonumber \\
 & \qquad \left(\nu^0 {+} \eta_m \nu^1\right)_{\mu \mu'}  \left(\tau^0 {+} \eta_t \tau^1\right)_{ss'} \left( T^{(\mu',-1)}_{(s',n)}(k)\right)
\bigg] \nonumber
\end{align}
here $n$ labels the two bands of the spin wave dispersion at
positive energies, and $T$ is the diagonalizing transformation
matrix, defined by $a = T \cdot \alpha$, such that
\begin{align}
a^\mu_{s,k} = \sum_{s',\mu'}  \left(T^{(\mu,\mu')}_{(s,s')}(k) \right) \alpha^{\mu'}_{s',k}\nonumber
\end{align}
where $\alpha$ are the Bogoliubov-diagonalized excitations, and
$\mu'=-1$ refers to the second component of the vector $\alpha$. The Bogoliubov transformation $T$ was computed numerically\cite{Gingras2004}. To compute the eigenvalues, it is sufficient to simply diagonalize $H(k) \nu^3$.

%\clearpage
%%%%%   BIBLIOGRAPHY   %%%%%%
\bibliography{Citations2016}

\end{document}